\documentclass[12pt]{iopart}
\pdfoutput=1
\usepackage[utf8]{inputenc}
\usepackage[english]{babel}
\usepackage[T1]{fontenc}
\expandafter\let\csname equation*\endcsname\relax % hack to work around amsmath conflict with the iopart style
\expandafter\let\csname endequation*\endcsname\relax
\usepackage{amsmath}
\usepackage{amsthm}
\usepackage{hyperref}
\usepackage{graphicx}
\usepackage[title,titletoc]{appendix}
\usepackage{tikz}
\usepackage{lipsum}
\usepackage{enumerate}
\usepackage{enumitem}
\usepackage{mathbbol}

\DeclareMathOperator*{\argmin}{argmin}

\newcommand{\R}{\mathbb{R}}
\newcommand{\C}{\mathbb{C}}

\newcommand{\EE}{\mathbb{E}}
\newcommand{\tp}{\tilde{\psi}}
\newcommand{\tw}{\tilde{W}}
\newcommand{\eps}{\epsilon}
\newcommand{\norm}[1]{\|{#1}\|}

\newcommand{\PP}[1]{\mathbb{P}\{{#1}\}}

\newcommand{\PPst}[2]{\mathbb{P}\{{#1}\  | \ {#2}\}}
\newcommand{\bignorm}[1]{\Big\|{#1}\Big\|}

\newcommand{\bigzero}{\mbox{\normalfont\normalsize\bfseries 0}}

\newcommand{\mf}{m^1}

\newcommand{\EEst}[2]{\mathbb{E}\left[{#1}\  \middle| \ {#2}\right]}

\newtheorem{theorem}{Theorem}
\newtheorem{lemma}{Lemma}
\newtheorem{remark}{Remark}
\newtheorem{assumption}{Assumption}

\newcommand{\reportnumber}{FERMILAB-PUB-22-842-QIS}
\usepackage[angle=0,scale=1,color=black,firstpage=true,opacity=1]{background}

\SetBgContents{\small\texttt{\reportnumber}}
\SetBgPosition{current page.north east}
\SetBgHshift{-3.5cm}
\SetBgVshift{-1.5cm}
\providecommand{\keywords}[1]
{
  \small	
  \textbf{\textit{Keywords---}} #1
}

\begin{document}

\title{Some aspects of noise in binary classification with quantum circuits
}
% Theoretical aspects of Krauss noise in binary classification using quantum circuits
% Some aspects of noise in binary classification using quantum circuits

\author{Yonghoon Lee$^1$, Do\~ga Murat K\"urk\c{c}\"uo\~glu$^{2}$, Gabriel N. Perdue$^{2}$}
\address{$^1$ The University of Chicago, IL, 60637, US}
\address{$^2$ Fermi National Accelerator Laboratory, Fermilab Quantum Institute, PO Box 500, Batavia, IL, 60510-0500, USA}
% Note -- I am dropping SQMS from our affiliation lists because this work was supported by QuantISED
%\address{$^3$ Fermi National Accelerator Laboratory, Superconducting Quantum Materials and Systems, PO Box 500, Batavia, IL, 60510-0500, USA}

%\maketitle

\begin{abstract}
We formally study the effects of a restricted single-qubit noise model inspired by real quantum hardware, and corruption in quantum training data, on the performance of binary classification using quantum circuits.
We find that, under the assumptions made in our noise model, that the measurement of a qubit is affected only by the noises on that qubit even in the presence of entanglement.
Furthermore, when fitting a binary classifier using a quantum dataset for training, we show that noise in the data can work as a regularizer, implying potential benefits from the noise in certain cases for machine learning problems.
%We illustrate practical examples of our findings with simulations. 
\end{abstract}


\section{Introduction}

In the last decade there have been numerous advances in quantum computing and machine learning.
Advancements in quantum hardware have allowed for substantial improvements in coherence times for qubits and gate fidelities and we are drawing near to being able to construct and operate practical quantum computers \cite{reagor2016quantum, siddiqi2021engineering, chen2014qubit}. 
Indeed, there have already been demonstrations quantum computers may offer advantages over other computing platforms for some problems \cite{Arute2019,morvan2023phase}. 
Despite recent progress, the qubits and gate operations are still strongly limited by noise, and decoherence effects are significant problems that must be addressed \cite{PhysRevX.10.041038}. 

Variational quantum classification (VQC) algorithms are among the most important approaches in quantum machine learning (QML) \cite{schuld2020circuit}. 
As with all quantum programs, the performance of VQCs is typically degraded when qubit and gate operations are imperfect. 
In this work, we investigate the effect of quantum noise in the binary classification task under a formally defined noise model.
While this noise model cannot perfectly describe the noise on a real quantum computer, it shares many of the features found in real hardware.
By choosing a formally defined noise model we are able to analyze it rigorously and provide foundations for future work under progressively more complex constructions of the noise.
The assumptions our noise model are a simplification of the operating conditions of a real device and we do not claim they fully capture all the important characteristics qubit noises.
Increasing the sophistication of the assumed noise model is a subject for future investigation.

Our findings can be summarized in three theorems. 
In Theorem 1, we find that under the assumption of only single-qubit noises in a quantum circuit, the measurement at a qubit is affected only by the error in that qubit.
This result holds even if the circuit includes entangling gates before the errors appear.
In Theorem 2, we derive a closed form formula for the corrupted measurement under Krauss and coherent errors. We find that quantum binary classifier is robust to such noise in the sense that the binary output remains the same unless the input value is sufficiently close to the classification boundary.
In Theorem 3, we study fitting a classifier using a quantum dataset, and find that noise in the data can function as a regularizer, implying that it can be beneficial in some cases.
Proofs of all three theorems are provided in the Appendix. 
% We additionally provide a few numerical experiments on simulated circuits using the Cirq\footnote{\url{https://quantumai.google/cirq}} library \cite{cirq_developers_2021} to support our findings. 
% \gabe{Our code may be found HERE --- link to repo, etc. --- Doga will fix}

\subsection{General notation}
In this section, we will explain the notation used in this work. 
For a given quantum state $\psi$, $\mathbb{P}_\psi\left\{m^{j} = 1\right\}$ denotes the probability to measure the qubit number $j$ in the excited state ($|1\rangle$) defined by $\psi$ in the computational basis. 
For example, we write $\mathbb{P}_{W\Phi(x)}\left\{m^{1} = 1\right\}$ to denote the probability of observing $|1\rangle$ from the first qubit of the output state when the input $x$ is encoded by $\Phi$ and then is input to a quantum circuit defined by $W$. 

\section{Effects of single qubit noises on the measurement of a quantum binary classifier}
\label{sec:singlequbitnoiseinmeas}

In this section, we discuss how our noise model (defined by arbitrary unitary single qubit noises) affects a variational quantum binary classifier.

\subsection{Problem setting}

Consider a quantum circuit semantically factorized into two pieces such that the input $x$ passes through an encoder $\Phi$ (to load the classical input) and then a parameterized circuit $W$.
We can apply such a quantum circuit for a binary classification task, where for simplicity we choose to base the classifier on the readout of a single qubit.
%\gabe{Is there any air here for people to complain about our result given we are only reading one qubit? --- if we used two, would it extend simply to noise on those two qubits, etc.?}

To make this idea concrete, suppose we have two classes $\{-1,1\}$, and for each $x$ in the input space $\mathcal{X}$, its class, denoted by $c(x)$, is either $-1$ or $1$. 
Now suppose we are given a quantum binary classifier $\hat{c}$, which classifies $x$ by
\begin{equation*}
\hat{c}(x) = \text{sign}\left(m(x)\right),
\end{equation*}
where
\begin{equation*}
    m(x) = \mathbb{P}_{W\Phi(x)}\{\mf = 1\} -\frac{1}{2} = \langle \Phi(x) | W^\dagger M_1 W | \Phi(x) \rangle  -\frac{1}{2}
\end{equation*}
denotes the probability of observing 1 at the first qubit. Here, $\mf$ denotes the measurement at the first qubit, and $M_1$ denotes the corresponding measurement operator, given by
\begin{equation*}
M_{1} =
\begin{bmatrix}
0 & 0 \\
0 & 1
\end{bmatrix}
\otimes I_2 \otimes \cdots I_2 = 
\begin{bmatrix}
\bigzero & \bigzero \\
\bigzero & I_{2^{n-1}}
\end{bmatrix}
.
\end{equation*}

%For example, $\hat{c}$ could be a classifier fit to some dataset --- in that case, $\hat{c}(x)$ works as an estimate of the true class of $x$.
%We have $\hat{c} = c$ in the settings where the true class $c$ is determined by the circuit.
%In any case, we aim to investigate the effect of noise in the output of a given quantum binary classifier.
%\gabe{Is the above paragraph necessary?}

Now we introduce noise model. 
We denote the corrupted circuit and encoder by $\tw$ and $\tilde{\Phi}$, respectively. We assume the following:
\begin{equation}\label{eqn:noise}
    \begin{split}
        \tw &= U W,\\
        \tilde{\Phi}(x) &= V \Phi(x),
    \end{split}
\end{equation}
where
\begin{align*}
    U &= U_1 \otimes U_2 \otimes \cdots \otimes U_n,\\
    V &= V_1 \otimes V_2 \otimes \cdots \otimes V_n,
\end{align*}
are unitaries with single qubit gates, which can be random.
In other words, our noise model assumes that an additional random gate appears for each of the qubits and no additional entanglement occurs through the noise (there are no two-qubit noises). 
For example, coherent error in a set of parameterized single-qubit rotation gates would satisfy this condition.

\subsection{Main results}

Our first finding is that the output distribution of the measurement at a qubit is not affected by the noises in the other qubits that occur after entangling gates in the circuit.
\begin{theorem}\label{thm:qubit_noise}
Assume the noise model defined in Equation ~\eqref{eqn:noise}, and let $U' = U_1 \otimes I_2 \otimes \cdots \otimes I_2$ and $V' = V_1 \otimes I_2 \otimes \cdots \otimes I_2$.
\begin{enumerate}
    \item If there's no encoder noise, i.e., $V \equiv I_{2^n}$,
    then the measurement of the first qubit does not depend on $\{U_k : 2 \leq k \leq n\}$. In other words,
\begin{equation*}
     \mathbb{P}_{UW\Phi(x)}\{\mf = 1\} =\mathbb{P}_{U'W\Phi(x)}\{\mf = 1\}
\end{equation*}
holds for any input $x$ and any set of random unitary matrices $U_2, U_3, \cdots ,U_n \in \C^{2 \times 2}$, where $\mf$ is a shorthand notation for the measurement of the first qubit in the circuit. 

\item Moreover, if $W$ has no entangling gates, i.e., it has the form of
\[W = W_1 \otimes W_2 \otimes \cdots \otimes W_n,\]
then
\begin{equation*}
     \mathbb{P}_{UWV\Phi(x)}\{\mf = 1\} =\mathbb{P}_{U'WV'\Phi(x)}\{\mf = 1\}
\end{equation*}
for any input $x$ and any set of random unitary matrices $U_2, U_3, \cdots ,U_n \in \C^{2 \times 2}$ and $V_2, V_3, \cdots ,V_n \in \C^{2 \times 2}$.
\end{enumerate}
\end{theorem}

\begin{remark}
In the proof of Theorem~\ref{thm:qubit_noise}, we actually prove a more general result --- the statements hold for $U = U_1 \otimes U_{2:n}$ and $V = V_1 \otimes V_{2:n}$ where $U_{2:n}$ and $V_{2:n}$ are any unitaries with $n-1$ qubits.
In other words, for the invariance of the distribution of the measurement at the first qubit, it is only required that there is no entangling noise that includes the first qubit.
\end{remark}

\begin{figure}
\includegraphics[width=0.45\textwidth]{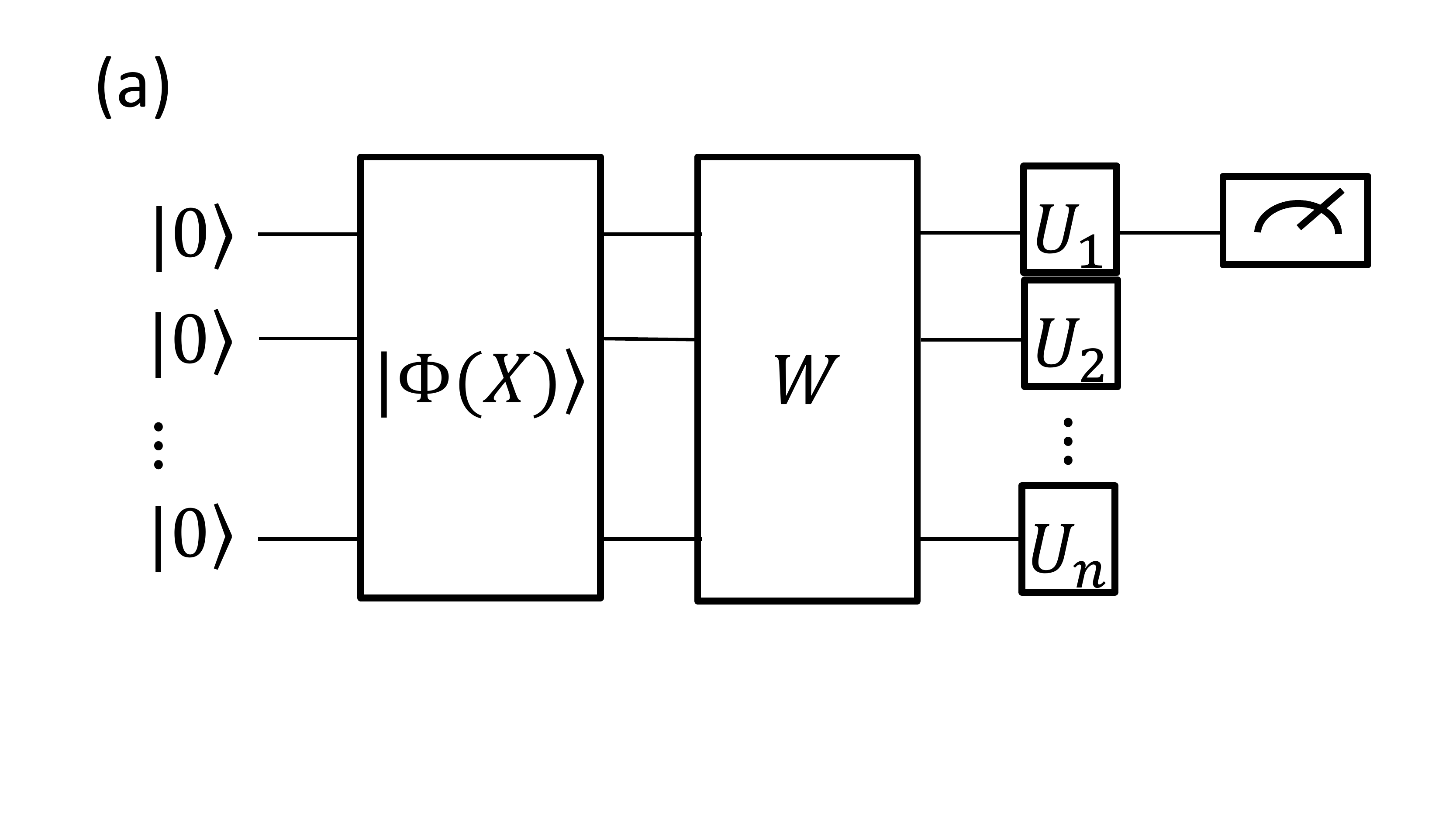} 
\includegraphics[width=0.45\textwidth]{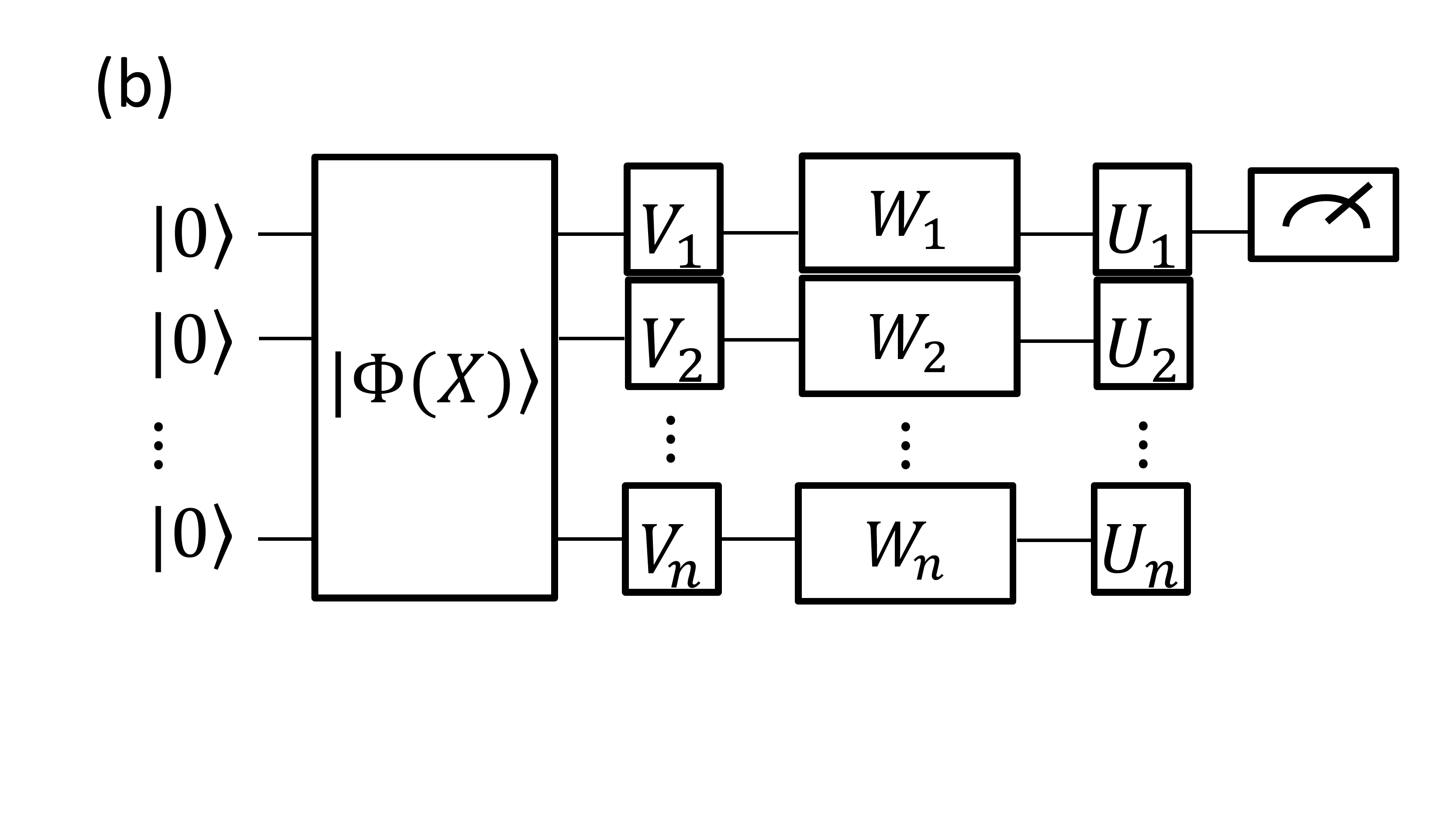} 
\caption{
The schematic for arrangement for Theorem~\ref{thm:qubit_noise}. 
The state $|\Phi(x)\rangle$ is prepared with some input $x$.
(a)
Then an arbitrary unitary $W$ is applied and then qubit $1$ is measured, under the presence of noise in $W$. Each $U_i$ is a random unitary which represents the noise applied to qubit $i$.
(b) 
The setting where $W$ has no entangling gates and we additionally have the noise $V = V_1 \otimes \cdots \otimes V_n$ in the encoder.
} 
\label{fig:fig1}
\end{figure}

Theorem~\ref{thm:qubit_noise} implies that no matter how many qubits we have, as long as we have only the single-qubit noises, the performance of the quantum classifier is affected only by the noise in the qubit we are measuring. In fact, this statement implies that we can simply remove all the single-qubit gates in the other qubits that appear after entanglements - it will change the quantum circuit, but the resulting classifier remains the same.

% In order to provide an empirical demonstration of this theorem, we constructed the circuits in Figure \ref{fig:fig1} (a) and (b) in a simulator model.
% In our example simulation, we ran a circuit as in Figure \ref{fig:fig1} with varying numbers of qubits (ranging from 4 to 16, counting by 2) with an error model that satisfies our assumed conditions.
% We showed the classifier performance was affected only by noise on the first (designated selector) qubit, as predicted by Theorem \ref{thm:qubit_noise}.
%We found zero difference between measurements of the first qubit without noise and measurements in the presence of noise as predicted by Theorem \ref{thm:qubit_noise}.
%and the error between the measurement of the first qubit without noise and the measurement in the presence of the noise is always not larger than the machine precision, $10^{-16}$.
%\gabe{The previous sentence is so unclear I am not sure how to fix it...}

%\gabe{Just to play Devil's Advocate here (for numerical experiments), if someone were skeptical of thrm 1, shouldn't we put noise on the other qubits also and then show the result is insensitive to the parameters for those noises --- e.g., you could make one of the other qubits so distorted by noise as to be random, but nothing changes?}  
% ---> this is what we did

%\gabe{This next block is confusing --- it feels like it is setting up the numerical experiments, but sort of isn't... what is going on between here and Theorem 2?}
%\gabe{Ok, I think I understand now --- we're just setting up Theorem 2?... this is written in a very confusing fashion.}
We now begin to prepare the ground for a new theorem.
We further consider the second clause of Theorem~\ref{thm:qubit_noise} and set up a specific instance for the individual noise terms to investigate the range of values that
\[\tilde{m}(x) = \mathbb{P}_{UWV\Phi(x)}\{\mf = 1\} - \frac{1}{2}\]
can have. 

For the noise in the encoder, it is sufficient to specify the model for $V_1$, which acts only on qubit 1, using the result of Theorem~\ref{thm:qubit_noise}:
\begin{equation}\label{eqn:bitflip_noise}
        \begin{split}
        &V_1 = \sigma_C, \text{ where }\\
        &\mathbb{P}\{C = j\} = 
        \begin{cases}
            1-3p &\text{ if } j = 0\\
            p &\text{ if } j = 1,2,3
        \end{cases}
        .
    \end{split}
\end{equation}
Here, we let $\sigma_0 = I_2$(i.e., no noise). In other words, we assume that an additional $\sigma_1$, $\sigma_2$, or $\sigma_3$ gate randomly appears, each with probability $p$, where $\sigma_j$ represents the standard Pauli matrices:
\begin{equation*}
    \sigma_1 = 
    \begin{bmatrix}
         0&1  \\
         1&0 
    \end{bmatrix}
    , 
    \sigma_2 = 
        \begin{bmatrix}
         0&-i  \\
         i&0 
    \end{bmatrix},
    \sigma_3 = 
    \begin{bmatrix}
         1&0  \\
         0&-1 
    \end{bmatrix}
    .
\end{equation*}
For the noise in $W$, again it is sufficient to set up a model for the noise in the first qubit only. We assume
\begin{equation}\label{eqn:rotation_noise}
    \begin{split}
        &U_1 = e^{-i \frac{\mu+\eps}{2}\sigma_{C'}}, \text{ where }\\
        &\mu \in [-\pi,\pi], \eps \sim N(0,\tau^2),
        \text{ and } \PP{C'=j} =
        \begin{cases}
            1-3q &\text{ if } j = 0\\
            q &\text{ if } j = 1,2,3.
        \end{cases}
        .
    \end{split}
\end{equation}
%\gabe{Choose a different jargon for variance vs Pauli gate --- Yonghoon will fix}
Here, $N(0,\tau^2)$ denotes the normal distribution with mean $0$ and variance $\tau^2$. Under this noise model, we may prove the following theorem:
\begin{theorem}
\label{thm:W_noise}
If $W$ has no entangling gate, then under the noise model defined by Equations~\eqref{eqn:bitflip_noise} and~\eqref{eqn:rotation_noise}, it holds that
\[\tilde{m}(x) \in \eta \cdot [m(x) - \delta, m(x) + \delta],\]
where
\begin{align*}
    \eta &= (1-4p)\cdot\left(1-2q(1-\cos \mu \cdot e^{-(1/2)\tau^2})\right),\\
    \delta &= \frac{2q \cdot |\sin \mu| \cdot e^{-(1/2)\tau^2}}{\eta}.
\end{align*}
\end{theorem}

Theorem~\ref{thm:W_noise} implies that as long as $p < 1/4$ and $|m(x)| > \delta$, the sign of $m(x)$ remains the same, i.e., $\text{sign}(\tilde{m}(x)) = \text{sign}(m(x))$. Note that $\lim_{\mu \rightarrow 0} \delta = \lim_{\tau \rightarrow 0} \delta = 0$, implying that the condition $|m(x)| > \delta$ tends to hold for moderate size of noise. In other words, the corrupted binary classifier tends to provide the same output as the original classifier, for $x$ values that are not too close to the separating boundary.

However, Theorem~\ref{thm:W_noise} also implies that the noise can shrink $m(x)$.
For example, in the setting where $\mu=0$, we have $\tilde{m}(x) = (1-4p)(1-2q(1-e^{-\frac{1}{2}\tau^2})) \cdot m(x)$. This shrinkage of $m(x)$ implies that a larger sample of measurements would be required for an accurate decision.

\section{Effect of single qubit noises in the training data on the fitted classifier}

In this section, we study problem of training a binary classifier on a noisy quantum dataset.

\subsection{Problem setting}

Suppose we have a ``native'' quantum dataset $(\psi_i,Y_i)_{1 \leq i \leq N} \subset \mathcal{H} \times \left\{-1,1\right\}$ where $\mathcal{H} \in \C^{2^n}$, and $\psi_i$'s are allowed to be entangled.
Examples of a quantum dataset include outputs from quantum sensors or a quantum computer.
The distinction between an encoded classical dataset and a naturally quantum dataset is not actually relevant here --- the point is we are not considering the encoding process if the dataset is classical.
Instead, we are beginning with quantum data that may be ``corrupted'' by noise.
The task is to fit a quantum circuit $W$ which classifies the points accurately. 

Here we consider empirical risk minimization to train the classifier. 
We use a quantum circuit parametrized by $\theta$, and minimize the risk $R(\theta)$, defined as
\begin{equation*}
   R(\theta) = \EE[\ell(m_\theta (\psi)\cdot Y)], 
\end{equation*}
where
\[ m_\theta(\psi) = \mathbb{P}_{W(\theta)\psi}\{\mf = 1\} - \frac{1}{2} = \langle \psi | W(\theta)^\dagger M_1 W(\theta) | \psi \rangle -\frac{1}{2}.\]
Here the expectation is taken with respect to the distribution of $(\psi,Y)$ (which we assume that each $(\psi_i,Y_i)$ follows), and $\ell$ denotes the loss function. The task is to fit a $W(\theta)$ of the form
\[W(\theta) = W_1(\theta_1) \times \cdots \times W_n(\theta_n),\]
that accurately classifies input $\psi$'s.
Note that we do not use entangling gates in the classifier.

In practice, we approximate the risk by empirical risk $\hat{R}_N(\theta)$:
\begin{equation*}
\hat{R}_N(\theta) = \frac{1}{N} \sum_{i=1}^N \ell(m_\theta(\psi_i) \cdot Y_i),
\end{equation*}
and the estimator of $\theta$ is obtained by minimizing:
\begin{equation*}
\hat{\theta}_N =  \argmin_{\theta \in \R^k} \hat{R}_N(\theta) = \argmin_{\theta \in \R^k} \frac{1}{N} \sum_{i=1}^N \ell(m_\theta(\psi_i)\cdot Y_i).
\end{equation*}

Now suppose we observe a noisy dataset $\{\tp_i\}_{1 \leq i \leq N} = \{V_i \psi_i\}_{1 \leq i \leq N}$ instead of $\left\{\psi_i\right\}_{1 \leq i \leq N}$. 
Then what we obtain by applying the above procedure is the corrupted empirical risk
\begin{equation*}
    \tilde{R}_N(\theta) = \frac{1}{N} \sum_{i=1}^N \ell(m_\theta(\tp_i)\cdot Y_i)
\end{equation*}
where
\begin{equation*}
    \tilde{m}_\theta(\psi_i) = \langle \psi_i | V_i W(\theta)^\dagger M_1 W(\theta) V_i | \psi_i \rangle -\frac{1}{2},
\end{equation*}
and the corrupted estimator
\begin{equation*}
\tilde{\theta}_N = \argmin_{\theta \in \R^k} \tilde{R}_N(\theta).
\end{equation*}

We will look into the performance of $\tilde{\theta}_N$ and compare it to the performance of the true estimator $\hat{\theta}_N$.

For the noise $V_i$, we consider the bitflip noise which we defined in Equation~\eqref{eqn:bitflip_noise}:

\begin{assumption}
\label{asm:noise}
The noise has the form of $V_i = V_i^1 \otimes \cdots \otimes V_i^n$, where
\[V_i^1 = \sigma_{C_i}, i=1,2,\cdots,N,\]
where $\{C_1,C_2,\cdots,C_N\}$ is an i.i.d sample from the distribution of $C$ defined as the following.
\begin{equation*}
    \mathbb{P}\{C = j\} = 
    \begin{cases}
            1-3p &\text{ if } j = 0\\
            p &\text{ if } j = 1,2,3
    \end{cases}
    .
\end{equation*}
\end{assumption}

We assume the following for the loss function $\ell$:

\begin{assumption}\label{asm:ell}
The loss function $\ell : (-\frac{1}{2},\frac{1}{2}) \mapsto [0,\infty)$ satisfies
\begin{enumerate}[label=(\alph*)]
    \item $\ell$ is convex and decreasing monotonically,
    \item $\ell(0)=1$.
\end{enumerate}
\end{assumption}

Assumption~\ref{asm:ell} is satisfied by most well known loss functions, such as hinge loss $\ell(t) = (1-t)_+$ and logistic loss $\log(1+\exp(-t))/\log(2)$.

\subsection{Main results}
Our main finding is that the noise in the data can work as a regularizer; hence classifier performance can benefit from the noise in some cases.
The idea follows the argument of~\cite{lee2022binary} which proves a similar result in the setting where we fit a classifier with classical data $\{X_i,Y_i\}_{1 \leq i \leq N}$.

\begin{theorem}\label{thm:regularization}
Under Assumptions~\ref{asm:noise} and~\ref{asm:ell}, the conditional expectation of the corrupted empirical risk given the training data $\{(\psi_i,Y_i)\}_{1 \leq i \leq N}$ is given by
\begin{equation*}
    \EE\Big[\tilde{R}_N(\theta) \,\Big|\, \psi_{1:N},Y_{1:N}\Big] = (1-4p) \cdot \left[\hat{R}_N(\theta) + \lambda\cdot \hat{P}_N(\theta)  \right],
\end{equation*}
where $\hat{P}_N(\theta)$ satisfies
\begin{equation*}
    \hat{P}_N(\theta) \geq \frac{1}{N}\sum_{i=1}^N \bigg(\frac{\ell\big(\frac{1}{3}|m_\theta(\psi_i)|  \big) + \ell\big( -\frac{1}{3} |m_\theta(\psi_i)| \big)}{2}\bigg).
\end{equation*}
and $\lambda = \frac{4p}{1-4p}$.
\end{theorem}

Therefore, the corrupted estimator $\tilde{\theta}_N$ can be thought as an approximation of the regularized estimator
\begin{equation}\label{eq:reg_loss}
    \theta_N^{regularized} = \argmin_{\theta \in \R^K} \left[ \hat{R}_N(\theta) + \lambda \cdot \hat{P}_N(\theta) \right]
\end{equation}
where $\lambda = 4p/(1-4p)$. Note that under Assumption~\ref{asm:ell}, $t \mapsto\frac{\ell(\frac{1}{3}t)+\ell(-\frac{1}{3}t)}{2}$ is nondecreasing at $t>0$, implying that having $\hat{P}_N(\theta)$ as a penalty term in~\eqref{eq:reg_loss} results in a shrinkage on $|m_\theta(\psi_i)|$ values. This prevents the resulting estimator $\tilde{\theta}_N$ from excessively overfitting the training data.

% \subsection{Numerical Experiments}

% ADD STUFF --- paragraph of interpretation for Figure for Theorem 3 when it shows up

% ADD Figure (code running)

% \gabe{Where is the stuff? :) --- Doga will fix... then Gabe and Yonghoon will read and edit}

\section{Conclusions}

In this work, we studied the effects a noise model defined by single-qubit errors on the performance of a binary classifier implemented with a quantum circuit. 

The most unexpected result from our finding is that even in the case of entangling gates in a quantum circuit, the noise on a gate or qubit is only going to affect the measurement from the specific qubit, i.e. the noise from other qubits is not going to corrupt the measurement at the qubit of interest. 

Our work also shows that the noise in the training data can even be beneficial for the goal of finding an optimal quantum classifier.
This is intuitive and has been anecdotally observed before, even on real quantum hardware \cite{Peters2021}.
But here we provide a formal argument supporting this phenomenon, although conditioned on a particular noise model.
These properties depend on the structure of the problem, e.g., binary classification with the measurement at the first qubit, which we discuss in this work.

Many questions yet remain.
Will we still observe such robustness of the measurement to the noise even in the case where we can have entangling gates in the noise model, or cross-talk?
These effects are known to be very important for real quantum computers, and, indeed, two qubit gate performance dominates concerns about single qubit gate performance on modern hardware platforms.
How will performance change in the setting where our goal is beyond binary classification?
We aim to explore these questions in future work.

\section{Acknowledgements}

This work was a project conducted while Y.L. was a student in the NSF MSGI and NSF Graduate Research Fellowship Program under Grant No. DGE-1644869.
D.K. and G. P. were supported for this work by the DOE/HEP QuantISED program grant ``HEP Machine Learning and Optimization Go Quantum,'' identification number 0000240323.

This document was prepared using the resources of the Fermi National Accelerator Laboratory (Fermilab), a U.S. Department of Energy (DOE), Office of Science, HEP User Facility. 
Fermilab is managed by Fermi Research Alliance, LLC (FRA), acting under Contract No. DE-AC02-07CH11359.

\onecolumn\newpage
\appendix

\bibliographystyle{plain}
\bibliography{references}

\section{Proof of Theorem~\ref{thm:qubit_noise}}

Let us write
\begin{equation*}
    U_{2:n} = U_2 \otimes \cdots \otimes U_n \text{ and } U_1 = 
    \begin{bmatrix}
        u_{11} & u_{12}\\
        u_{21} & u_{22}
    \end{bmatrix}
    ,
\end{equation*}
so that $U$ can be written as
\begin{align*}
    U = U_1 \otimes U_{2:n} = 
    \begin{bmatrix}
        u_{11} \cdot U_{2:n} & u_{12} \cdot U_{2:n}\\
        u_{21} \cdot U_{2:n} & u_{22} \cdot U_{2:n}
    \end{bmatrix}
    ,
\end{align*}
and write
\begin{align*}
    \Phi(x) = 
            \begin{bmatrix}
                \Phi_1(x) \\
                \Phi_2(x)
            \end{bmatrix}.
\end{align*}
Then we have
\begin{align*}
    &\mathbb{P}_{UW\Phi(x)}\left\{\mf = 1\right\}  \\
    = &\EE\left[\mathbb{P}_{UW\Phi(x)}\left\{\mf = 1 \ | \ U \right\}\right]\\
    = &\EE\big[\langle \Phi(x) | W^\dagger U^\dagger M_1 U W  |\Phi(x)\rangle\big]\\
    = &\EE\Bigg[\Bigg\|
         \begin{bmatrix}
            \bigzero & \bigzero \\
            \bigzero & I_{2^{n-1}}
        \end{bmatrix}
    \begin{bmatrix}
        u_{11} \cdot U_{2:n} & u_{12} \cdot U_{2:n}\\
        u_{21} \cdot U_{2:n} & u_{22} \cdot U_{2:n}
    \end{bmatrix}
        \begin{bmatrix}
            \Phi_1(x)\\
            \Phi_2(x)
        \end{bmatrix}
        \Bigg\|^2\Bigg]\\
    = &\EE\left[\bignorm{u_{21} \cdot U_{2:n} \Phi_1(x) + u_{22} \cdot U_{2:n} \Phi_2(x)}^2\right]\\
    = &\EE\left[\bignorm{u_{21} \cdot \Phi_1(x) + u_{22} \cdot \Phi_2(x)}^2\right], \textnormal{\quad since $U_{2:n}$ is unitary}
\end{align*}
which does not depend on the distribution of $U_{2:n}$. Hence, the first claim is proved.

Similarly, for the second claim we write
\begin{equation*}
    W_{2:n} = W_2 \otimes \cdots \otimes W_n\text{ and } W_1 = 
    \begin{bmatrix}
        w_{11} & w_{12}\\
        w_{21} & w_{22}
    \end{bmatrix}
    ,
\end{equation*}
and
\begin{equation*}
    V_{2:n} = V_2 \otimes \cdots \otimes V_n \text{ and } V_1 = 
    \begin{bmatrix}
        v_{11} & v_{12}\\
        v_{21} & v_{22}
    \end{bmatrix}
    ,
\end{equation*}
then it holds that
\begin{equation*}
    \begin{split}
        &\mathbb{P}_{UWV\Phi(x)}\left\{\mf = 1\right\}\\
        = &\EE\left[\mathbb{P}_{UWV\Phi(x)}\left\{\mf = 1 \ | \ U, V \right\}\right]\\
        = &\EE\big[\langle \Phi(x) | V^\dagger W^\dagger U^\dagger M_1 U W V |\Phi(x)\rangle\big]\\
        = &\EE\Bigg[\Bigg\|
         \begin{bmatrix}
            \bigzero & \bigzero \\
            \bigzero & I_{2^{n-1}}
        \end{bmatrix}
        \begin{bmatrix}
            u_{11} \cdot U_{2:n} & u_{12} \cdot U_{2:n}\\
            u_{21} \cdot U_{2:n} & u_{22} \cdot U_{2:n}
        \end{bmatrix}
        \begin{bmatrix}
            w_{11} \cdot W_{2:n} & w_{12} \cdot W_{2:n}\\
            w_{21} \cdot W_{2:n} & w_{22} \cdot W_{2:n}
        \end{bmatrix}
        \begin{bmatrix}
            v_{11} \cdot V_{2:n} & v_{12} \cdot V_{2:n}\\
            v_{21} \cdot V_{2:n} & v_{22} \cdot V_{2:n}
        \end{bmatrix}
        \cdot
        \begin{bmatrix}
            \Phi_1(x)\\
            \Phi_2(x)
        \end{bmatrix}
        \Bigg\|^2\Bigg] \\
        = &\EE\bigg[\Big\|(u_{21}w_{11}v_{11}+u_{22}w_{21}v_{11}+u_{21}w_{12}v_{21}+u_{22}w_{22}v_{21}) \cdot U_{2:n}W_{2:n}V_{2:n} \Phi_1(x)\\
        &\qquad\qquad + (u_{21}w_{11}v_{12}+u_{22}w_{21}v_{12}+u_{21}w_{12}v_{22}+u_{22}w_{22}v_{22}) \cdot U_{2:n}W_{2:n}V_{2:n} \Phi_2(x)\Big\|^2\bigg]\\
        = &\EE\bigg[\Big\|(u_{21}w_{11}v_{11}+u_{22}w_{21}v_{11}+u_{21}w_{12}v_{21}+u_{22}w_{22}v_{21})\Phi_1(x)\\
        &\qquad\qquad + (u_{21}w_{11}v_{12}+u_{22}w_{21}v_{12}+u_{21}w_{12}v_{22}+u_{22}w_{22}v_{22}) \Phi_2(x)\Big\|^2\bigg],
    \end{split}
\end{equation*}
where the last equality holds since $U_{2:n}W_{2:n}V_{2:n}$ is unitary. Therefore, $\tilde{m}_\theta(x)$ does not depend on $U_{2:n}$, $W_{2:n}$ and $V_{2:n}$ and thus the second claim is proved.

\section{Proof of Theorem~\ref{thm:W_noise}}

We first introduce a lemma. 
%\gabe{Yonghoon will review proof of theorem 2}

\begin{lemma}\label{lem:m_theta_j}
Assume noise model~\eqref{eqn:noise},~\eqref{eqn:bitflip_noise}, and,~\eqref{eqn:rotation_noise}, and suppose $W$ has the form of
\[W = W_1 \otimes \cdots \otimes W_n.\]
Define
\[m^{jl}(x) = \mathbb{P}_{UWV\Phi(x)}\left\{\mf = 1 \,|\, C=j, C'=l \right\} - \frac{1}{2}\]
for $j \in \left\{0,1,2,3\right\}$ and $l \in \left\{0,1,2,3\right\}$. Then the following statements hold.
\begin{enumerate}[label=(\alph*)] 
    \item $m^{0l}(x) + m^{1l}(x) + m^{2l}(x) + m^{3l}(x) = 0$ for all $l \in \{0,1,2,3\}$.
    \item $m^{00}(x) = m^{03}(x) = m(x)$, and $m^{01}(x) = m^{02}(x)$ with
    \[|m^{01}(x) - \cos \mu \cdot e^{-\frac{1}{2}\tau^2}\cdot m(x)| \leq \frac{1}{2}|\sin \mu|\cdot e^{-\frac{1}{2}\tau^2}.\]
\end{enumerate}
\end{lemma}

Applying (a) of Lemma~\ref{lem:m_theta_j}, we have

\begin{align*}
    \PPst{m^1=1}{C'=l} &= \EEst{\PPst{m^1=1}{C, C'=l}}{C'=l}\\
    &= (1-3p)\cdot m^{0l}(x) + p \cdot m^{1l}(x) + p \cdot m^{1l}(x) + p \cdot m^{1l}(x)\\
    &= (1-4p)\cdot m^{0l}(x) + p \cdot \left(m^{0l}(x) + m^{1l}(x) + m^{2l}(x) + m^{3l}(x)\right)\\
    &= (1-4p)\cdot m^{0l}(x).
\end{align*}
It follows that
\begin{align*}
    \tilde{m}(x) &= \EE\left[\PPst{m^1=1}{C'=l}\right]\\
    &= (1-3q)\cdot (1-4p)\cdot m^{00}(x) + q \cdot (1-4p)\cdot m^{01}(x) + q \cdot (1-4p)\cdot m^{02}(x) + q \cdot (1-4p)\cdot m^{03}(x)\\
    &= (1-4p)\cdot\left[(1-2q) \cdot m(x) + 2q \cdot m^{01}(x)\right]\\
    &= (1-4p)\cdot\left[(1-2q) \cdot m(x) + 2q \cdot \left(u(x) + \cos \mu \cdot e^{-\frac{1}{2}\tau^2}\cdot m(x)\right)\right]\\
    &\;\;\;\;\, (\text{where } u(x) = m^{01}(x) - \cos \mu \cdot e^{-\frac{1}{2}\tau^2}\cdot m(x))\\
    &= (1-4p)\cdot \left(1-2q + \cos \mu \cdot e^{-\frac{1}{2}\tau^2}\right) \cdot m(x) + (1-4p)\cdot 2q \cdot u(x). 
\end{align*}
The desired inequality follows from $|u(x)| \leq \frac{1}{2}|\sin \mu|\cdot e^{-\frac{1}{2}\tau^2}$ which holds by (b) of Lemma~\ref{lem:m_theta_j}.

\section{Proof of Theorem~\ref{thm:regularization}}

We apply the following lemma:

\begin{lemma}\label{lem:reg}
Under Assumption~\ref{asm:ell}, the following inequality holds for any $a,b,c,d \in \R$ with $a+b+c+d=0$.
\begin{equation*}
    \frac{\ell(a)+\ell(b)+\ell(c)+\ell(d)}{4} \geq \frac{\ell(\frac{1}{3}|a|) + \ell(-\frac{1}{3}|a|)}{2}.
\end{equation*}
\end{lemma}

The proof is given in the next section. Now we prove Theorem~\ref{thm:regularization}. The arguments apply the idea of \cite{lee2022binary}. We first compute
\begin{align*}
&\EE\Big[\tilde{R}_N(\theta) \,\Big|\, \psi_{1:N},Y_{1:N}\Big]\\
= & \EE\Big[\EE\Big[\tilde{R}_N(\theta) \,\Big|\, \psi_{1:N},Y_{1:N}, C\Big]\Big]\\
= &(1-3p) \cdot \frac{1}{N}\sum_{i=1}^N \ell(m_\theta^0(\psi_i)\cdot Y_i) + p \cdot \frac{1}{N}\sum_{i=1}^N \ell(m_\theta^1(\psi_i)\cdot Y_i)+ p \cdot \frac{1}{N}\sum_{i=1}^N \ell(m_\theta^2(\psi_i)\cdot Y_i) + p\cdot \frac{1}{N}\sum_{i=1}^N \ell(m_\theta^3(\psi_i)\cdot Y_i)\\
= &(1-4p)\hat{R}_N(\theta) + 4p \cdot \frac{1}{N}\sum_{i=1}^N \frac{\ell(m_\theta(\psi_i)\cdot Y_i)+\ell(m_\theta^{1}(\psi_i)\cdot Y_i)+\ell(m_\theta^{2}(\psi_i)\cdot Y_i)+\ell(m_\theta^{3}(\psi_i)\cdot Y_i)}{4}
\end{align*}
where
\[m_\theta^j(x) = \langle \Phi(x) | (\sigma_j \otimes I_{2^{n-1}}) W(\theta)^\dagger M_1 W(\theta) (\sigma_j \otimes I_{2^{n-1}}) | \Phi(x) \rangle - \frac{1}{2}\]
Applying Lemma~\ref{lem:m_theta_j} with $U = I_{2^n}$, we have
\[m_\theta(x)+m_\theta^1(x)+m_\theta^2(x)+m_\theta^3(x) = 0\]
Therefore, by lemma~\ref{lem:reg}, it holds that
\begin{multline*}
    \frac{1}{N}\sum_{i=1}^N \frac{\ell(m_\theta(\psi_i)\cdot Y_i)+\ell(m_\theta^{1}(\psi_i)\cdot Y_i)+\ell(m_\theta^{2}(\psi_i)\cdot Y_i)+\ell(m_\theta^{3}(\psi_i)\cdot Y_i)}{4}\\
    \geq \frac{1}{N}\sum_{i=1}^N \bigg(\frac{\ell\big(\frac{1}{3}|m_\theta(\psi_i)|  \big) + \ell\big( -\frac{1}{3} |m_\theta(\psi_i)| \big)}{2}\bigg),
\end{multline*}
and this proves the claim.

\section{Proof of lemmas}

\subsection{Proof of Lemma~\ref{lem:m_theta_j}}
By Theorem~\ref{thm:qubit_noise}, we may assume $U = U_1 \otimes I_{2^{n-1}}$ and $V = V_1 \otimes I_{2^{n-1}}$. Fix any $l$, and compute
\begin{equation*}
    \begin{split}
         m^{jl}(x) &= \PPst{\mf = 1}{C = j, C'=l} - \frac{1}{2}\\
         &= \EE\big[\PPst{\mf = 1}{C=j, C'=l, \eps}\big] - \frac{1}{2}\\
         &= \EE\left[\langle \Phi(x) | (\sigma_j \otimes I_{2^{n-1}})^\dagger W^\dagger (U_1 \otimes I_{2^{n-1}})^\dagger M_1 (U_1 \otimes I_{2^{n-1}}) W (\sigma_j \otimes I_{2^{n-1}}) |\Phi(x)\rangle\right] -\frac{1}{2}.
    \end{split}
\end{equation*}
Next, we write 
\begin{align*}
    U_1 = 
    \begin{bmatrix}
        u_{11} & u_{12}\\
        u_{21} & u_{22}
    \end{bmatrix}
    , W_1 =
    \begin{bmatrix}
        w_{11} & w_{12}\\
        w_{21} & w_{22}
    \end{bmatrix}
\end{align*}
and write $\Phi(x)$ in a block matrix form.
\begin{equation*}
    \begin{split}
        \Phi(x) = 
        \begin{bmatrix}
            \Phi_1(x)\\ \Phi_2(x)
        \end{bmatrix}
        ,\quad \Phi_1(x),\Phi_2(x) \in \mathcal{C}^{2^{n-1}}.
    \end{split}
\end{equation*}
Now if $j=0$, we have
\begin{equation*}
    \begin{split}
        m^{0l}(x) &= \EE\Bigg[\Bigg\|
        \begin{bmatrix}
            \bigzero & \bigzero \\
            \bigzero & I_{2^{n-1}}
        \end{bmatrix}
        \begin{bmatrix}
            u_{11} I_{2^{n-1}} & u_{12} I_{2^{n-1}}\\
            u_{21} I_{2^{n-1}} & u_{22} I_{2^{n-1}}
        \end{bmatrix}
        \begin{bmatrix}
            w_{11} W_{2:n} & w_{12} W_{2:n}\\
            w_{21} W_{2:n} & w_{22} W_{2:n}
        \end{bmatrix}
        \begin{bmatrix}
            \Phi_1(x)\\
            \Phi_2(x)
        \end{bmatrix}
        \Bigg\|^2\Bigg] - \frac{1}{2}\\
        &= \EE\Big[\norm{(u_{21}w_{11}+u_{22}w_{21}) W_{2:n} \Phi_1(x) + (u_{21}w_{12}+u_{22}w_{22}) W_{2:n}\Phi_2(x)}^2\Big] - \frac{1}{2}\\
        &= \EE\Big[\norm{(u_{21}w_{11}+u_{22}w_{21}) \Phi_1(x) + (u_{21}w_{12}+u_{22}w_{22}) \Phi_2(x)}^2\Big] - \frac{1}{2}.
    \end{split}
\end{equation*}
If j=1, we have
\begin{equation*}
    \begin{split}
        m^{1l}(x) &= \EE\Bigg[\Bigg\|
        \begin{bmatrix}
            \bigzero & \bigzero \\
            \bigzero & I_{2^{n-1}}
        \end{bmatrix}
        \begin{bmatrix}
            u_{11} I_{2^{n-1}} & u_{12} I_{2^{n-1}}\\
            u_{21} I_{2^{n-1}} & u_{22} I_{2^{n-1}}
        \end{bmatrix}
        \begin{bmatrix}
            w_{11} W_{2:n} & w_{12} W_{2:n}\\
            w_{21} W_{2:n} & w_{22} W_{2:n}
        \end{bmatrix}
        \begin{bmatrix}
            \bigzero & I_{2^{n-1}}\\
            I_{2^{n-1}} & \bigzero
        \end{bmatrix}
        \begin{bmatrix}
            \Phi_1(x)\\
            \Phi_2(x)
        \end{bmatrix}
        \Bigg\|^2\Bigg] - \frac{1}{2}\\
        &= \EE\Big[\norm{(u_{21}w_{11}+u_{22}w_{21}) W_{2:n} \Phi_2(x) + (u_{21}w_{12}+u_{22}w_{22}) W_{2:n}\Phi_1(x)}^2\Big] - \frac{1}{2}\\
        &= \EE\Big[\norm{(u_{21}w_{11}+u_{22}w_{21}) \Phi_2(x) + (u_{21}w_{12}+u_{22}w_{22}) \Phi_1(x)}^2\Big] - \frac{1}{2}.
    \end{split}
\end{equation*}
By similar procedure, we obtain
\[m^{2l}(x) = \EE\Big[\norm{-(u_{21}w_{11}+u_{22}w_{21}) \Phi_2(x) + (u_{21}w_{12}+u_{22}w_{22}) \Phi_1(x)}^2\Big] - \frac{1}{2}\]
and
\[m^{3l}(x) = \EE\Big[\norm{(u_{21}w_{11}+u_{22}w_{21}) \Phi_1(x) - (u_{21}w_{12}+u_{22}w_{22}) \Phi_2(x)}^2\Big] - \frac{1}{2}.\]
Now we define $a_1 = u_{21}w_{11}+u_{22}w_{21}$ and $a_2 = u_{21}w_{12}+u_{22}w_{22}$, then we can write
\begin{align*}
    &m^0(x) + m^1(x) + m^2(x) + m^3(x)\\
    = &\EE\Big[\norm{a_1 \Phi_1(x) + a_2 \Phi_2(x)}^2 +\norm{a_1 \Phi_2(x) + a_2 \Phi_1(x)}^2 + \norm{-a_1 \Phi_2(x) + a_1 \Phi_1(x)}^2 + \norm{a_1 \Phi_1(x) - a_2 \Phi_2(x)}^2\Big] -2\\
    = &\EE\Big[2(|a_1|^2+|a_2|^2)(\norm{\Phi_1(x)}^2+\norm{\Phi_2(x)}^2)\Big] - 2\\
    = &2\cdot 1 \cdot \norm{\Phi(x)}^2 - 2\\
    = &0,
\end{align*}
where the third equality holds since the vector $(a_1,a_2)$ is the second row of the matrix $U_1W_1$ which is unitary. Hence, (a) is proved.

Next, define
\begin{align*}
    \Psi(x) = 
    \begin{bmatrix}
        \Psi_1(x)\\
        \Psi_2(x)
    \end{bmatrix}
    = W | \Phi(x) \rangle
\end{align*}
and compute
\begin{equation*}
    \begin{split}
        m^{00}(x) &=\EE\big[\PPst{\text{measurement} = 1}{\eps, C = 0, C'=0}\big] - \frac{1}{2}\\
        &=\EE\big[\langle \Phi(x) | W^\dagger U^\dagger M_1 U W |\Phi(x)\rangle\big] - \frac{1}{2}\\
        &=\EE\Bigg[\Bigg\|\left(\cos \frac{\mu+\eps}{2} - i\sin \frac{\mu+\eps}{2}\right)\cdot
         \begin{bmatrix}
            \bigzero & \bigzero \\
            \bigzero & I_{2^{n-1}}
        \end{bmatrix}
        \begin{bmatrix}
            \Psi_1(x)\\
            \Psi_2(x)
        \end{bmatrix}
        \Bigg\|^2\Bigg] - \frac{1}{2}\\
        &= \norm{\Psi_2(x)}^2-\frac{1}{2} (= m(x) ),
    \end{split}
\end{equation*}
and
\begin{equation*}
    \begin{split}
        m^{01}(x) &=\EE\big[\PPst{\text{measurement} = 1}{\eps, C = 0, C'=1}\big] - \frac{1}{2}\\
        &=\EE\Bigg[\Bigg\|
         \begin{bmatrix}
            \bigzero & \bigzero \\
            \bigzero & I_{2^{n-1}}
        \end{bmatrix}
        \begin{bmatrix}
            \cos \frac{\mu_1+\eps_1}{2} \cdot I_{2^{n-1}}  & -i \sin\frac{\mu_1+\eps_1}{2} \cdot I_{2^{n-1}}\\
            -i \sin\frac{\mu_1+\eps_1}{2} \cdot I_{2^{n-1}}  & \cos \frac{\mu_1+\eps_1}{2} \cdot I_{2^{n-1}} 
        \end{bmatrix}
        \begin{bmatrix}
            \Psi_1(x)\\
            \Psi_2(x)
        \end{bmatrix}
        \Bigg\|^2\Bigg] - \frac{1}{2}\\
        &= \EE\left[\bignorm{-i \sin\frac{\mu_1+\eps_1}{2} \cdot \Psi_1(x) + \cos \frac{\mu_1+\eps_1}{2} \cdot \Psi_2(x)}^2\right] - \frac{1}{2}\\
        &= \EE\left[\frac{1}{2} + \cos (\mu_1+\eps_1) \cdot \left(\norm{\Psi_2(x)}^2-\frac{1}{2}\right) - \sin (\mu_1+\eps_1) \cdot \text{Re}(i\Psi_1(x)^\dagger \Psi_2(x))\right] - \frac{1}{2}\\
        &=\cos \mu_1 \cdot m(x) \cdot e^{-\frac{1}{2}\tau^2} + \sin \mu_1 \cdot e^{-\frac{1}{2}\tau^2} \cdot \text{Im}(\Psi_1(x)^\dagger \Psi_2(x)).
    \end{split}
\end{equation*}
By similar computations we have
\begin{align*}
    m^{03}(x) &= \norm{\Psi_2(x)}^2-\frac{1}{2} = m^{00}(x)\\
    m^{02}(x) &= \cos \mu_1 \cdot m(x) \cdot e^{-\frac{1}{2}\tau^2} + \sin \mu_1 \cdot e^{-\frac{1}{2}\tau^2} \cdot \text{Im}(\Psi_1(x)^\dagger \Psi_2(x)) = m^{01}(x).
\end{align*}
Note that
\begin{equation*}
    |\text{Im}(\Psi_1(x)^\dagger \Psi_2(x))| \leq |\Psi_1(x)^\dagger \Psi_2(x)| \leq \frac{|\Psi_1(x)|^2+|\Psi_2(x)|^2}{2} = \frac{1}{2}.
\end{equation*}
This proves (b).

\subsection{Proof of Lemma~\ref{lem:reg}}
Let us define
\[ s = \frac{1}{2}\cdot \max\left\{|a+b|,|a+c|,|a+d|\right\}.\]
Then we have
\[\frac{\ell(a)+\ell(b)+\ell(c)+\ell(d)}{4} \geq \frac{\ell(s) + \ell(-s)}{2},\]
since
\begin{align*}
    \frac{\ell(a)+\ell(b)+\ell(c)+\ell(d)}{4} &\geq \frac{1}{2}\cdot \left[\ell\left(\frac{a+b}{2}\right)+\ell\left(\frac{c+d}{2}\right)\right]\textnormal{\quad by Jensen's inequality}\\
    &= \frac{1}{2}\cdot \left[\ell\left(\frac{a+b}{2}\right)+\ell\left(-\frac{a+b}{2}\right)\right]\textnormal{\quad since $a+b+c+d=0$}\\
    &= \frac{1}{2}\cdot \left[\ell\left(\frac{|a+b|}{2}\right)+\ell\left(-\frac{|a+b|}{2}\right)\right]
\end{align*}
and the above procedure also holds when $(a+b,c+d)$ is replace by $(a+c,b+d)$ or $(a+d,b+c)$. Note that $t \mapsto \frac{\ell(t)+\ell(-t)}{2}$ is nondecreasing at $t > 0$ because of the convexity of $\ell$. Now by definition of $s$, we have
\[ |a| = |b+c+d| = \frac{1}{2}|(b+c)+(c+d)+(d+c)| = \frac{1}{2}|(a+b)+(a+c)+(a+d)| \leq 3 s,\]
and therefore
\[ \frac{\ell(s) + \ell(-s)}{2} \geq \frac{\ell(\frac{1}{3}|a|) + \ell(-\frac{1}{3}|a|)}{2}. \]
The desired inequality immediately follows from these observations.

\end{document}